\documentclass[11pt,a4paper]{article}

\usepackage[total={6.5in, 10.2in}, top=0.85in, left=0.8in, right=0.8in,marginparwidth=0.8in, footskip=0.0in, marginparsep=0.5cm]{geometry}
\usepackage{lastpage,fancyhdr,graphicx}
\usepackage{epstopdf}
\usepackage{subcaption}

\usepackage[T1]{fontenc}
\usepackage[english]{babel}

\usepackage{layouts}

\usepackage{nameref,hyperref}
\usepackage[bbgreekl]{mathbbol}
\usepackage{etoolbox}
\usepackage{mathtools}
\usepackage{csquotes}
\usepackage[right]{lineno}
\usepackage{lmodern}
\usepackage{tgheros}
\usepackage{threeparttable}
\usepackage{microtype}
\DisableLigatures[f]{encoding = *, family = * }
\usepackage{esint}
\usepackage{capt-of}
\usepackage[capitalise]{cleveref}
\usepackage{algorithm2e}

\usepackage{changepage}

\usepackage[aboveskip=7pt,labelfont=bf,labelsep=space,singlelinecheck=off, font={small,sf}]{caption}

\usepackage[backend=bibtex,
    natbib=true,
    style=numeric-comp,
    sorting=none,
    maxbibnames=12,
    maxcitenames=2,
    giveninits=true,
    date=year,
    doi=true,
    url=true, isbn=false, eprint=false]{biblatex}
\DeclareFieldFormat[article]{pages}{#1}
\DeclareFieldFormat[inproceedings]{pages}{#1}
\makeatletter
\renewcommand{\@biblabel}[1]{\quad#1.}
\makeatother

\usepackage{color}
\usepackage{graphicx}
\usepackage{booktabs}

\usepackage[table]{xcolor}
\definecolor{Gray}{RGB}{100,100,100}
\definecolor{LightGray}{RGB}{245,245,249}
\definecolor{naturelight}{RGB}{248,231,215}
\definecolor{naturedark}{RGB}{204,139,136}

\definecolor{myblue}{RGB}{0,100,200}
\definecolor{myred}{RGB}{204,102,0}
\definecolor{mygreen}{RGB}{0,200,50}
\definecolor{cottoncandy}{RGB}{254,200,216}
\definecolor{tumblue}{RGB}{0,100,189}
\colorlet{lightblue}{cyan!100!white}

\usepackage{color,soul}

\setul{0.5ex}{0.3ex}
\setulcolor{lightblue}

\usepackage{chemformula}

\usepackage{marginnote}
\reversemarginpar

\hypersetup{
    colorlinks,
    linkcolor=myred,
    citecolor=myblue,
    urlcolor=myblue
}

\usepackage{amsthm}
\usepackage{enumitem}

\newtheorem*{remark}{Remark}

\usepackage{amsmath}
\usepackage{siunitx}
\usepackage[scr=boondoxo]{mathalfa}

\usepackage{listings}
\DeclarePairedDelimiterX{\infdivx}[2]{}{}{%
{#1}\;\big\|\;{#2}%
}

\newcommand{\mat}[1]{\boldsymbol{#1}}
\newcommand{\vct}[1]{\boldsymbol{#1}}

\newcommand{\inv}{^{-1}}

\newcommand{\NP}{^{\textsf{NP}}}
\newcommand{\magn}{_{\textsf{mag}}}

\newcommand{\phiNP}{\phi\NP}

\newcommand*\dd{\mathop{}\!\mathrm{d}}

\newcommand{\be}{\begin{equation}}
        \newcommand{\ee}{\end{equation}}
\newcommand{\bc}{\begin{center}}
        \newcommand{\ec}{\end{center}}
\newcommand{\bd}{\begin{description}}
        \newcommand{\ed}{\end{description}}
\newcommand{\bi}{\begin{itemize}}
        \newcommand{\ei}{\end{itemize}}

\definecolor{codebackground}{rgb}{0.95, 0.95, 0.92}
\usepackage{setspace}
\usepackage{tabularx}
\usepackage{sansmath}
\usepackage{mhchem}

\DeclareCaptionFormat{mylst}{\hrule#1#2#3}

\usepackage{newfloat}
\DeclareFloatingEnvironment[fileext=frm,placement={tbp},name=Algorithm]{myfloat}

\usepackage{tikz}
\usetikzlibrary{shapes,snakes}
\newcommand*\circled[1]{\tikz[baseline=(char.base)]{
        \node[shape=circle,draw,inner sep=1pt] (char) {#1};}}

\usepackage{nomencl}
\makenomenclature

\usepackage{etoolbox}
\renewcommand\nomgroup[1]{%
    \item[\bfseries
                \ifstrequal{#1}{A}{Dimensions and corresponding indices}{%
                    \ifstrequal{#1}{C}{Sobol method related symbols}{%
                        \ifstrequal{#1}{B}{Random quantities}{%
                            \ifstrequal{#1}{D}{Metamodel related symbols}{%
                                \ifstrequal{#1}{O}{Other symbols}{}}}}}%
          ]}

\usepackage{csquotes}
\begin{document}
\lstdefinestyle{interfaces}{
float=bp,
floatplacement=bp,
abovecaptionskip=-5pt
}

{
\sffamily
\begin{flushleft}
    {\LARGE
        An efficient computational model of the in-flow capturing of magnetic nanoparticles by a cylindrical magnet for cancer nanomedicine
    }
    \newline
    \\
    Barbara Wirthl\textsuperscript{1,*},
    Vitaly Wirthl\textsuperscript{2},
    Wolfgang A. Wall\textsuperscript{1}
    \\
    \bigskip
    {
        \small
        \textbf{1} Institute for Computational Mechanics, Technical University of Munich, TUM School of Engineering and Design, Department of Engineering Physics \& Computation, Garching b. München, Germany.
        \\
        \textbf{2} Max Planck Institute of Quantum Optics (MPQ), Garching b. München, Germany.
        \\
        \bigskip
        * B. Wirthl, E-mail: \href{barbara.wirthl@tum.de}{barbara.wirthl@tum.de}
    }

\end{flushleft}

{
\setlength{\parindent}{0cm}
{\large \textbf{Abstract} \smallskip}

Magnetic nanoparticles have emerged as a promising approach to improving cancer treatment.
However, many novel nanoparticle designs fail in clinical trials
due to a lack of understanding of how to overcome the \textit{in vivo} transport barriers.
To address this shortcoming, we develop a novel computational model aimed at the study of magnetic nanoparticles \textit{in vitro} and \textit{in vivo}. 
In this paper, we present an important building block for this overall goal, namely an efficient computational model of the in-flow capture of magnetic nanoparticles by a cylindrical permanent magnet in an idealised test setup.
We use a continuum approach based on the Smoluchowski advection-diffusion equation, combined with a simple approach to consider the capture at an impenetrable boundary, and derive an analytical expression for the magnetic force of a cylindrical magnet of finite length on the nanoparticles.
This provides a simple and numerically efficient way to study different magnet configurations and their influence on the nanoparticle distribution in three dimensions.
Such an \textit{in silico} model can increase insight into the underlying physics, help to design novel prototypes and serve as a precursor to more complex systems \textit{in vivo} and \textit{in silico}.
}
}
\section{Introduction}

Over the last three decades, nanoparticles have emerged as a promising approach to improve the effectiveness of cancer treatment because of their potential for sophisticated functionalisation and ability to accumulate in tumours \cite{Shi2017}.
Magnetic nanoparticles are of particular interest because of their ability to be controlled by an external magnetic field.
To capture the drug-loaded magnetic nanoparticles in the target region, the applied magnetic force has to be strong enough to overcome fluid forces due to the blood flow or the interstitial fluid flow and further transport barriers, e.g., the extracellular matrix, the blood vessel wall, which the nanoparticles must cross, and different interfaces.
However, this is often hard to achieve because of the inherently weak magnetic forces produced by an applied magnetic field---especially deeper in the body \cite{Hewlin2023}.
Because of those (and other) challenges, the design and successful application of magnetic nanoparticle-based cancer therapy is very demanding and almost hopeless purely via trial-and-error approaches in experimental research. 
Here, computational models can help by predicting the distribution of nanoparticles depending on the applied magnetic field and guide the design of novel prototypes.

\begin{figure}[tbp]
    \centering
    \includegraphics[width=135mm]{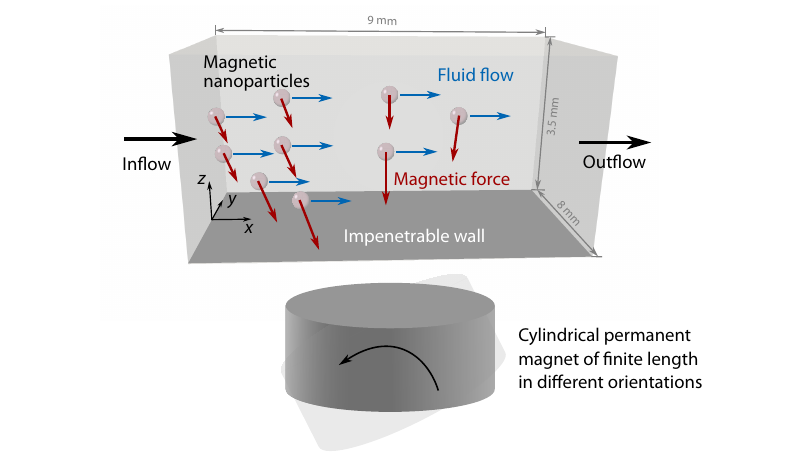}
    \caption{Idealised test setup. %
        The magnetic nanoparticles are dispersed in the fluid flowing through the channel. %
        A cylindrical permanent magnet is placed below the channel and exerts a magnetic force on the magnetic nanoparticles to capture them at the bottom wall, which is impenetrable. %
    }
    \label{fig:IdealisedSetup}
\end{figure}

On the way towards a comprehensive computational model of the capture of magnetic nanoparticles, we here start with an idealised test setup, illustrated in \cref{fig:IdealisedSetup}:
a cylindrical permanent magnet is placed below a channel to capture the magnetic nanoparticles dispersed in the fluid flowing through the channel.
This setup, even though simplified, contains the essential physics of the capture of magnetic nanoparticles:
the magnetic force exerted by the magnet combined with the fluid flow, which is known to be a major transport barrier \textit{in vivo} \cite{Stillman2020}.
The bottom wall of the domain is impenetrable, so the captured nanoparticles accumulate at the wall.
Such an idealised test setup, including tumour spheroids in the microfluidic channel, is used in experimental research, e.g., \cite{Nguyen2020, Behr2022, Kappes2022}, because it allows insight into the fundamental physics of the capture of magnetic nanoparticles and serves as a precursor to more complex \textit{in vivo} systems.
The approaches and results of the current work are essential for exactly modelling the experimental setup where tumour spheroids are placed in the microfluidic channel \cite{Wirthl2023c}. 

To model the transport of magnetic nanoparticles, two approaches are the most common in the literature \cite{Hallmark2019}:
the first approach models the nanoparticles as discrete particles, while the second approach assumes that the nanoparticles behave as a continuum ferrofluid.
The first approach considers the different forces acting on each particle individually, and Newton's second law then describes the movement of each particle \cite{Cregg2009, Cregg2010, Palovics2020, Palovics2022}.
This allows investigating the aggregation of the nanoparticles and the formation of particle clusters, e.g., chains \cite{Palovics2020, Palovics2022}.
Nevertheless, when the system has a size of millimetres or centimetres, the number of particles in the domain is on the order of $10^9$, which limits the applicability of this approach.
Moreover, we are not interested in the movement of each particle individually.
The second approach builds on the assumption that the nanoparticles have an infinitely strong coupling with the base fluid, and this fluid-particle mixture is described as a whole by the classical fluid equations, e.g., the Navier--Stokes equations \cite{Li2007, Li2008, Munir2009, Khashan2011}.
Hence, the magnetic force is part of the momentum balance equation.
This approach, however, does not allow the nanoparticles to move relative to the fluid \cite{Khashan2011}.

To overcome the limitations of both approaches, we take a different approach here, similar to \cite{Furlani2006}:
we model the nanoparticles in a continuum sense, but consider that the nanoparticles can move relative to the advecting fluid due to diffusion and the exerted magnetic force.
We therefore use an advection-diffusion equation to model the concentration of the nanoparticles and include the magnetic force directly in this equation.
In this contribution, we address two specific challenges:
the boundary condition at an impenetrable boundary and the efficient evaluation of the magnetic force exerted by a cylindrical magnet of finite length.

Concerning the first challenge (the boundary condition at an impenetrable boundary), \citeauthor{Khashan2011}~\cite{Khashan2011} state that most contributions in the literature that study the transport of magnetic nanoparticles in a continuum sense do not consider the boundary condition at the impenetrable boundary, which results in nanoparticles leaving the domain through the boundary.
This however would be questionable for our long term goals, as we among others also want to be able to model complex time-dependent scenarios, where we should not loose nanoparticles.
Hence, we present a simple approach to model the capture of magnetic nanoparticles at an impenetrable boundary.

Concerning the second challenge (the efficient evaluation of the magnetic force), we derive an analytical expression for the magnetic force on magnetic nanoparticles exerted by a cylindrical magnet of finite length.
This presents a huge advantage of our approach, as it allows us to directly evaluate the magnetic force on the nanoparticles with minimal computational effort compared to numerically solving Maxwell's equations.
At the same time, we can study different magnet orientations in three dimensions---in contrast to two-dimensional models, commonly used in the literature \cite{Furlani2006, Furlani2007, Etgar2010, Shaw2018, Yeo2021, Hewlin2023}, which have the severe limitation that they assume the magnet to be infinitely long and oriented perpendicular to the two-dimensional domain.

In the following, we first introduce the equations for the advection-diffusion problem, including an impenetrable boundary, in \cref{Sec:SmoluchowskiAdvectionDiffusionEquation}.
We then present the analytical expression for the magnetic force on the nanoparticles in \cref{Sec:MagneticForceOnTheNanoparticles}.
\cref{Sec:NumericalExamples} presents and discusses numerical examples for both and \cref{Sec:Conclusion} draws a conclusion.

\section{Methods}

\subsection{Transport of nanoparticles: Smoluchowski advection-diffusion equation}
\label{Sec:SmoluchowskiAdvectionDiffusionEquation}

We assume that the magnetic nanoparticles are dispersed in the fluid in a stable colloidal suspension, so the particle concentration $\phiNP$ is small ($\phiNP \ll 1$).
We therefore assume that nanoparticles do not interact with each other, i.e., we neglect the inter-particle forces for now, and we assume that the nanoparticles do not influence the fluid flow.

The concentration of the nanoparticles $\phiNP$ is governed by the mass balance equation
\begin{equation}
    \frac{\partial \phiNP}{\partial t}
    +
    \vct{\nabla} \cdot \vct{q}_{\textsf{tot}}
    =
    0,
\end{equation}
where $\vct{q}_{\textsf{tot}}$ denotes the total flux.
We here assume that neither sources nor sinks are present.
The total flux is the sum of three contributions
\begin{equation}
    \vct{q}_{\textsf{tot}}
    =
    \vct{q}_{\textsf{diff}}
    +
    \vct{q}_{\textsf{adv}}
    +
    \vct{q}\magn,
    \label{Eq:TotalFlux}
\end{equation}
arising from diffusion, advection and the magnetic force, respectively.

First, the diffusive flux $\vct{q}_{\textsf{diff}}$ arises from a local concentration gradient $\vct{\nabla} \phiNP$ and is described by Fick's first law
\begin{equation}
    \vct{q}_{\textsf{diff}}
    =
    -D \vct{\nabla} \phiNP
\end{equation}
with the diffusion coefficient $D$.
Second, the advective flux $\vct{q}_{\textsf{adv}}$ arises from the velocity $\vct{v}_{\textsf{adv}}$ of the fluid advecting the nanoparticles and is given by
\begin{equation*}
    \vct{q}_{\textsf{adv}}
    =
    \vct{v}_{\textsf{adv}}\phiNP.
\end{equation*}
The fluid velocity $\vct{v}_{\textsf{adv}}$ might be given by solving the underlying flow problem, e.g., the Navier--Stokes equations or Darcy flow in a porous medium.
For simplicity of presentations in this paper, we here directly prescribe the velocity of the fluid.

When the nanoparticles are not only subjected to fluid flow but additionally to a magnetic force, we include an additional magnetophoretic flux $\vct{q}\magn$ depending on the magnetic force $\vct{F}\magn$.
Typically, the resulting magnetophoretic velocity $\vct{v}\magn$ is assumed to be directly proportional to the applied force, e.g., see \cite{Takayasu1983, Furlani2006}, resulting in a magnetophoretic flux of
\begin{equation}
    \vct{q}\magn
    =
    \vct{v}\magn\,\phiNP
    \quad\text{with}\quad
    \vct{v}\magn
    =
    \zeta\inv \vct{F}\magn,
    \label{Eq:MagneticVelocityScalarMobility}
\end{equation}
where $\zeta = 6 \pi \mu^\ell R\NP$ is the mobility of a particle of radius $R\NP$ in a fluid with dynamic viscosity\footnotemark{} $\mu^\ell$, based on Stokes' law.
Altogether, this results in the Smoluchowski advection-diffusion equation \cite{Smoluchowski1915}
\begin{equation}
    \frac{\partial \phiNP}{\partial t}
    -
    \vct{\nabla} \cdot \left( D \vct{\nabla} \phiNP \right)
    +
    \vct{\nabla} \cdot \left( \vct{v}_{\textsf{adv}} \phiNP \right)
    +
    \vct{\nabla} \cdot \left( \zeta\inv \vct{F}\magn \, \phiNP \right)
    =
    0.
\end{equation}
\footnotetext{$\mu^\ell$ denotes the dynamic viscosity of the fluid by the superscript $\ell$ (\textit{liquid}) to distinguish it from the magnetic vacuum permeability $\mu_0$ introduced in \cref{Sec:MagneticForceOnTheNanoparticles}.}

\cref{Eq:MagneticVelocityScalarMobility} assumes that the velocity is always directly proportional to the applied force.
Now consider an example of a channel with an impenetrable wall and a force perpendicular to the wall, as sketched in \cref{fig:IdealisedSetup}:
the force results in a velocity perpendicular to the wall, which in turn results in the nanoparticles leaving the domain through the impenetrable wall---which is obviously not physical.

We therefore introduce the mobility tensor $\vct{\mathcal{M}}$, which relates the magnetophoretic velocity to the magnetic forces, i.e.,
\begin{equation}
    \vct{v}\magn
    =
    \vct{\mathcal{M}} \vct{F}\magn,
\end{equation}
or more explicitly,
\begin{equation}
    \begin{bmatrix}
        v_x \\ v_y \\ v_z
    \end{bmatrix}
    =
    \begin{bmatrix}
        \mathcal{M}_{xx} & \mathcal{M}_{xy} & \mathcal{M}_{xz} \\
        \mathcal{M}_{yx} & \mathcal{M}_{yy} & \mathcal{M}_{yz} \\
        \mathcal{M}_{zx} & \mathcal{M}_{yz} & \mathcal{M}_{zz} \\
    \end{bmatrix}
    \begin{bmatrix}
        F_x \\ F_y \\ F_z
    \end{bmatrix},
\end{equation}
similar to \cite{Pimponi2014, Chen2020}.
The mobility is a tensor field $\vct{\mathcal{M}}(\vct{x})$ that depends on the position $\vct{x}$.
Now, a force in a specific direction does not necessarily result in a velocity in that direction, but only if the particle can move in this direction, i.e., if the mobility is non-zero.
At an impenetrable wall, the nanoparticles cannot move in the direction perpendicular to the wall, i.e., into the wall, and thus the mobility is zero in this direction, resulting also in zero velocity.
All off-diagonal entries of the mobility tensor $\vct{\mathcal{M}}$ are also zero:
a force in one direction only causes a velocity in the same direction and no \enquote{shear} velocity.
Inside the domain $\mathcal{M}_{xx} = \mathcal{M}_{yy} = \mathcal{M}_{zz} = \zeta\inv$, which reduces back to \cref{Eq:MagneticVelocityScalarMobility}.
At the impenetrable wall (at $z = 0$), the diagonal entries tangential to the wall still equal the scalar mobility, i.e., $\mathcal{M}_{xx} = \mathcal{M}_{yy} = \zeta\inv$.
However, the entry perpendicular to the wall is zero $\mathcal{M}_{zz} = 0$, as already mentioned above.
The mobility tensor at the impenetrable wall is thus given by
\begin{equation}
    \vct{\mathcal{M}}_{\textsf{wall}}
    =
    \begin{bmatrix}
        \zeta\inv & 0         & 0 \\
        0         & \zeta\inv & 0 \\
        0         & 0         & 0 \\
    \end{bmatrix}.
\end{equation}
The key point here is that we employ the mobility as a tensor field---as opposed to a scalar.

The final form of the Smoluchowski advection-diffusion equation is then given by
\begin{equation}
    \frac{\partial \phiNP}{\partial t}
    -
    \mat{\nabla} \cdot (D \mat{\nabla} \phiNP)
    +
    \mat{\nabla} \cdot ( \vct{v}_{\textsf{adv}} \phiNP)
    +
    \mat{\nabla} \cdot \left(\vct{\mathcal{M}} \, \vct{F}\magn \phiNP \, \right)
    =
    0.
    \label{Eq:SmoluchowskiAdvectionDiffusionEquation}
\end{equation}

To solve this equation in space and time, we employ the standard Galerkin procedure to obtain the weak form of the equation and then discretise the equation in space using the finite element method (FEM) and in time using the backward Euler method.
We use our in-house parallel multiphysics research code BACI \cite{BACI2023} as a computational framework.

\begin{remark}[Stabilisation]
    In our case, \cref{Eq:SmoluchowskiAdvectionDiffusionEquation} is dominated by the two convective terms, which causes numerical instabilities when using the standard Galerkin procedure.
    We therefore use the streamline upwind Petrov--Galerkin (SUPG) method \cite{Brooks1982} to stabilise the equation, where numerical diffusion along streamlines is introduced in a consistent manner \cite{John2006}.
    We choose the stabilisation parameter $\tau$ based on \citeauthor{Codina2002}~\cite{Codina2002}.
\end{remark}

\subsection{Magnetic force on the nanoparticles}
\label{Sec:MagneticForceOnTheNanoparticles}

Due to the permanent magnet, the magnetic nanoparticles are subjected to a static non-homogenous external magnetic field $\vct{H}$ leading to a force $\vct{F}\magn$.
This force however does not only depend on the magnetic field but also the magnetic response of the particles.

Due to the small size of the particles, we assume that they can be modelled as an equivalent point dipole located at the centre of the particle (\emph{effective dipole moment approach} \cite{Jones1995, Hallmark2019, Furlani2006}).
Also, due to the small size, the nanoparticles are superparamagnetic:
they are magnetised with a large magnetic susceptibility $\chi\NP$ when an external magnetic field is applied but do not retain their magnetisation after the external magnetic field is removed.
Hence, when a superparamagnetic nanoparticle is placed in an external magnetic field, it magnetises, resulting in a magnetic moment $\vct{m}\NP$.
The force on the magnetic dipole induced in the nanoparticle is then given by
\begin{equation}
    \vct{F}\magn
    =
    \mu_0
    \left(\vct{m}\NP \cdot \vct{\nabla} \right) \vct{H},
\end{equation}
with the magnetic vacuum permeability $\mu_0$\footnotemark{}.
\footnotetext{%
    We assume that the fluid (water) and air are non-magnetic, and thus assume that their permeability is equal to the vacuum magnetic permeability defined as $\mu_0 = \SI{1.25663706212(19)e-6}{\newton\per\ampere\squared}$ \cite{Tiesinga2021}.
    The exact values for water and air differ from the value for vacuum at the fifth and seventh decimal place, respectively.
}
Using the magnetisation $\vct{M}\NP$ as the magnetic moment per volume, i.e., $\vct{M}\NP = \vct{m}\NP / V\NP$ with $V\NP$ being the volume of the nanoparticle, the force can be written as
\begin{equation}
    \vct{F}\magn
    =
    \mu_0
    V\NP
    \left(\vct{M}\NP \cdot \vct{\nabla} \right) \vct{H}.
\end{equation}
Thus, the force depends on the magnetisation of the nanoparticle and the derivatives of the applied magnetic field.

The magnetised nanoparticles also produce a magnetic field, affecting the nearby nanoparticles.
For now, we assume that the magnetic force that the nanoparticles exert on each other is negligible compared to the magnetic force exerted by the external magnetic field---which is a valid assumption for low concentrations of nanoparticles and hence large distances between the nanoparticles \cite{Khashan2011, Keaveny2008, Han2010, Furlani2006, Woinska2013}.
We will investigate and discuss the validity of this assumption in \cref{sec:InterParticleForces}.

\begin{figure}[tb]
    \centering
    \includegraphics[width=135mm]{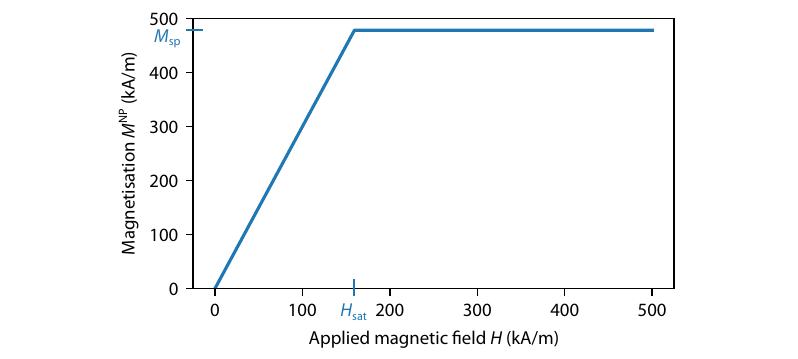}
    \caption{Magnetisation curve for a superparamagnetic nanoparticle with linear magnetisation and saturation above an applied magnetic field of $H_{\textsf{sat}}$, assuming a saturation magnetisation of $M_{\textsf{sp}} = \SI{478}{\kilo\ampere\per\metre}$ \cite{Furlani2006} and a magnetic susceptibility of $\chi\NP \gg 1$ \cite{Furlani2006, Sun2008, McNamara2017}.}
    \label{fig:MagnetisationCurve}
\end{figure}

\paragraph{Magnetisation model}
To relate the magnetisation of the nanoparticle to the applied magnetic field, we use a linear magnetisation model with saturation, given by
\begin{equation}
    \vct{M}\NP = f(|\vct{H}|) \vct{H}
    \quad \text{with} \quad
    f(|\vct{H}|) =
    \begin{cases*}
        \; \frac{3\chi\NP}{3 + \chi\NP}   & if $|\vct{H}| < H_{\textsf{sat}}$    \\
        \frac{M_{\textsf{sp}}}{|\vct{H}|} & if $|\vct{H}| \geq H_{\textsf{sat}}$
    \end{cases*}
    \label{Eq:DemagFactorHallmark}
\end{equation}
with $M_{\textsf{sp}}$ being the saturation magnetisation and $H_{\textsf{sat}}$ the field strength for which the particle reaches saturation, as presented by \cite{Furlani2006, Hallmark2019, Takayasu1983}.
An example of such a magnetisation curve is shown in \cref{fig:MagnetisationCurve}.
If the particle is below saturation, its magnetisation is proportional to the applied magnetic field
\begin{equation}
    \vct{M}\NP = \frac{3\chi\NP}{3 + \chi\NP} \vct{H},
\end{equation}
and the particle reaches saturation for
\begin{equation}
    H_{\textsf{sat}} = \frac{\chi\NP + 3}{3 \chi\NP} M_{\textsf{sp}},
\end{equation}
which can be derived based on the effective dipole moment approach \cite{Jones1995, Furlani2006}.
Above saturation, the magnetisation is equal to the saturation magnetisation $M_{\textsf{sp}}$
\begin{equation}
    \vct{M}\NP = M_{\textsf{sp}} \frac{\vct{H}}{|\vct{H}|}.
\end{equation}
The magnetisation is always aligned with the applied magnetic field.

Finally, as discussed above, the particles are superparamagnetic: their magnetic susceptibility is much higher than the magnetic susceptibility of paramagnetic materials, i.e., $\chi\NP \gg 1$ \cite{Furlani2006, Sun2008, McNamara2017}.
\cref{Eq:DemagFactorHallmark} can then be simplified to
\begin{equation}
    f(|\vct{H}|)
    =
    \begin{cases*}
        \; 3                              & if $|\vct{H}| < \frac{1}{3}M_{\textsf{sp}}$    \\
        \frac{M_{\textsf{sp}}}{|\vct{H}|} & if $|\vct{H}| \geq \frac{1}{3}M_{\textsf{sp}}$
    \end{cases*}.
    \label{Eq:DemagFactorFurlani}
\end{equation}

In sum, the magnetic force on the nanoparticles is given by
\begin{equation}
    \vct{F}\magn
    =
    \mu_0
    V\NP
    f(|\vct{H}|)
    \left(\vct{H} \cdot \vct{\nabla} \right) \vct{H},
    \label{Eq:MagneticForceFinal}
\end{equation}
which shows that the magnetic force depends both on the strength of the magnetic field and its derivatives.

\paragraph{Analytical expression for the magnetic field}

Usually, the magnetic field $\vct{H}$ is obtained by solving Maxwell's equations numerically.
Analytic expressions are only well-known for some classic textbook cases:
the magnetic field of point multipoles and infinitely long wires carrying a current \cite{Jackson2021}.
However, for a finite-length cylindrical magnet, which we have here, \citeauthor{Derby2010}~\cite{Derby2010} and \citeauthor{Caciagli2018}~\cite{Caciagli2018} presented analytic expressions based on the elliptic integrals.
These analytic expressions are beneficial because the magnetic quantities can be evaluated at all coordinates with minimal computational effort compared to numerically solving Maxwell's equations, e.g., using the FEM.

In the following, we summarise the analytic expression for the magnetic field, as presented by \cite{Derby2010, Caciagli2018}, and then extend this by deriving the analytic expressions for the magnetic force.
The cylindrical magnet is magnetised in the longitudinal direction.
The field components of the magnetic field $\vct{H}$ in cylindrical coordinates $(\rho, \phi, z)$ are given by
\begin{align}
    H_\rho(\rho, z)
     & =
    \frac{M_s R\magn}{\pi} \left[
        \alpha_+ P_1(k_+) - \alpha_- P_1(k_-)
        \right] \label{Eq:Hrho} \\
    H_z(\rho, z)
     & =
    \frac{M_s R\magn}{\pi (\rho + R\magn)} \left[
        \beta_+ P_2(k_+) - \beta_- P_2(k_-)
        \right] \label{Eq:Hz}
\end{align}
and $H_\phi = 0$ due to the radial symmetry of the system \cite{Derby2010, Caciagli2018}\footnotemark{}.
Here, $M_s$ denotes the magnetisation of the cylindrical magnet and $R\magn$ its radius.
The origin of the cylindrical coordinate system is located at the centre of the magnet.
The two auxiliary functions $P_1$ and $P_2$ are defined as
\begin{align}
    P_1(k)
     & =
    \mathcal{K}\left(1 - k^2\right)
    -
    \frac{2}{1 - k^2} \left[\mathcal{K}\left(1 - k^2\right) - \mathcal{E}\left(1 - k^2\right) \right] \label{Eq:P1} \\
    P_2(k)
     & =
    -\frac{\gamma}{1 - \gamma^2} \left[\mathcal{\Pi}\left(1 - \gamma^2, 1 - k^2\right) - \mathcal{K}\left(1 - k^2\right) \right]
    -
    \frac{1}{1 - \gamma^2}
    \left[\gamma^2 \mathcal{\Pi}\left(1 - \gamma^2, 1 - k^2\right) - \mathcal{K}\left(1 - k^2\right) \right] \label{Eq:P2}
\end{align}
with the following auxiliary variables
\begin{equation*}
    \rho_{\pm} = R\magn \pm \rho, \quad
    \zeta_{\pm} = \frac{L\magn}{2} \pm z, \quad
    \alpha_{\pm} = \frac{1}{\sqrt{ \zeta_{\pm}^2 + \rho_{+}^2}}, \quad
    \beta_{\pm} = \zeta_{\pm} \alpha_{\pm}, \quad
    \gamma = -\frac{\rho_-}{\rho_+}, \quad
    k_{\pm} = \sqrt{\frac{\zeta_{\pm}^2 + \rho_{-}^2}{\zeta_{\pm}^2 + \rho_{+}^2}}
\end{equation*} %
and $L\magn$ being the length of the cylindrical magnet.
\footnotetext{%
    \cref{Eq:Hrho,Eq:Hz} are mathematically well-behaved except on the edge of the magnet at $\rho = \pm R\magn$ and $z = \pm \frac{L\magn}{2}$ \cite{Derby2010}.%
}
\cref{Eq:P1,Eq:P2} are based on the complete elliptic integrals of the first, second and third kind, which in Legendre's notation are written as
\begin{align}
    \mathcal{K}(m)     & = \int\limits_{0}^{\pi/2} \frac{\dd \theta}{\sqrt{1 - m \sin^2 \theta}}                       \label{Eq:EllipticKm}  \\
    \mathcal{E}(m)     & = \int\limits_{0}^{\pi/2} \sqrt{1 - m \sin^2 \theta} \dd \theta                              \label{Eq:EllipticEm}   \\
    \mathcal{\Pi}(n,m) & = \int\limits_{0}^{\pi/2} \frac{\dd \theta}{(1 - n \sin^2 \theta)\sqrt{1 - m \sin^2 \theta}}. \label{Eq:EllipticPim}
\end{align}
All three kinds of elliptic integrals can be efficiently evaluated using Carlson's functions $R_F$, $F_D$ and $R_J$ \cite{Carlson1979, Carlson1981} as
\begin{align}
    \mathcal{K}(m)     & = R_F(0, 1 - m, 1)           \phantom{\frac{m}{3}}      \\
    \mathcal{E}(m)     & = R_F(0, 1 - m, 1) - \frac{m}{3} R_D(0, 1 - m, 1)       \\
    \mathcal{\Pi}(n,m) & = R_F(0, 1 - m, 1) + \frac{n}{3} R_J(0, 1 - m, 1 - n) .
\end{align}
\textit{Numerical Recipies}~\cite{Press2007} provides algorithms and source code for evaluating Carlson's functions, which are also implemented in Mathematica \cite{Mathematica} and SciPy \cite{SciPy}.
\begin{remark}[Parameter and sign conventions in the elliptic integrals]
    Note that \textit{Numerical Recipies} \cite[p. 315]{Press2007} uses a different sign convention for the variable $n$ in the third elliptic integral, such that
    \begin{equation}
        \mathcal{\Pi}(n,m) = \int\limits_{0}^{\pi/2} \frac{\dd \theta}{(1 + n \sin^2 \theta)\sqrt{1 - m \sin^2 \theta}} = R_F(0, 1 - m, 1) - \frac{n}{3} R_J(0, 1 - m, 1 + n).
    \end{equation}
    Additionally, \cite{Caciagli2018} use the convention with parameter $k$, where $m = \tilde{k}^2 = \sqrt{1 - k^2}$ in their Eq.~(6) in \cite{Caciagli2018}.
    Mathematica \cite{Mathematica} and SciPy \cite{SciPy} however use the parameter $m$, as presented here in \cref{Eq:EllipticKm,Eq:EllipticEm,Eq:EllipticPim}.
\end{remark}

\begin{figure}[tbp]
    \centering
    \includegraphics[width=\linewidth]{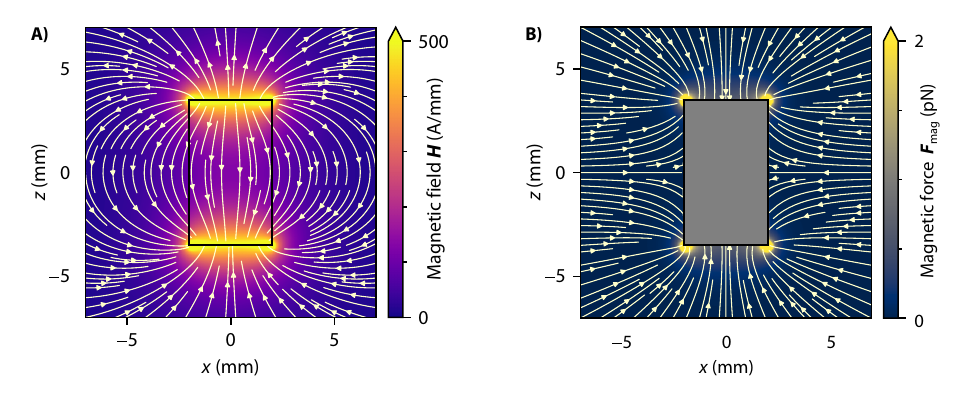}
    \caption{Magnetic field $\vct{H}$ (A) and magnetic force $\vct{F}\magn$ (B) on the nanoparticles of a cylindrical magnet with radius $R\magn = \SI{2}{\milli\meter}$ and length $L\magn = \SI{7}{\milli\meter}$.%
    }
    \label{fig:ExampleMagnet}
\end{figure}

\cref{fig:ExampleMagnet}A shows an example of the magnetic field $\vct{H}$ of a cylindrical magnet with radius $R\magn = \SI{2}{\milli\meter}$, length $L\magn = \SI{7}{\milli\meter}$ and magnetisation $M_s = \SI{1e6}{\ampere\per\meter}$.
Inside the magnet, the magnetic field is given by $\vct{H} = \frac{\vct{B}}{\mu_0} - \vct{M}_s$, with the magnetic flux density $\vct{B}$.
For a longitudinally magnetised magnet, the magnetisation vector is $\vct{M}_s = M_s \vct{e}_z$, with $\vct{e}_z$ being the unit vector in $z$-direction.
The magnetisation vector is constant inside and zero outside the magnet.
The result for the magnetic field $\vct{H}$ in \cref{fig:ExampleMagnet}A is qualitatively well known:
the magnetic field lines start at one pole and end at the other, forming fanned-out circular segments around the magnet.

\paragraph{Analytical expression for the magnetic force}
As discussed above, the magnetic force $\vct{F}\magn$ depends on the magnetic field and its derivatives.
Since the first derivatives of the elliptic integrals are known analytically, we can derive an analytical expression for the magnetic force $\vct{F}\magn$.
Evaluating \cref{Eq:MagneticForceFinal} for the analytical expression for the magnetic field results in the force components given by
\begin{equation}
    \begin{split}
        F_\rho(\rho, z) = &\;
        \frac{\mu_0 V\NP f(|\vct{H}|) M_s^2}{4  \pi^2  \rho_{+} \rho^3 a_1 a_2  a_3  a_4}
        \Biggl[
            \left(
            \frac{a_4 c_1 \zeta_{+} \mathcal{E}(\psi_{+})}{\alpha_{-}}
            +\frac{a_3 c_2 \zeta_{-} \mathcal{E}(\psi_{-})}{\alpha_{+}}
            - \frac{a_3 a_4 \zeta_{+} \mathcal{K}(\psi_{+})}{\alpha_{-}}
            - \frac{a_3 a_4 \zeta_{-} \mathcal{K}(\psi_{-})}{\alpha_{+}}
            \right)
            \rho^2 Q_2 \\
            &+
            \left(
            \frac{a_4 (b_1^2 + b_3 \rho^2) \mathcal{E}(\psi_{+})}{\alpha_{-}}
            - \frac{a_3 (b_2^2 + b_4 \rho^2) \mathcal{E}(\psi_{-})}{\alpha_{+}}
            + \frac{a_3 a_4 b_2 \mathcal{K}(\psi_{-})}{\alpha_{+}}
            - \frac{a_3 a_4 b_1  \mathcal{K}(\psi_{+})}{\alpha_{-}}
            \right)
            \rho_{+} Q_1
            \Biggr]
    \end{split}
    \label{Eq:FrhoLongitudinal}
\end{equation}
and
\begin{equation}
    \begin{split}
        F_z (\rho, z) = &\;
        \frac{\mu_0 V\NP f(|\vct{H}|) M_s^2}{4 \pi^2 a_1 a_2 a_3 a_4}
        \Biggl[
            \left(
            \frac{a_3 a_4 \zeta_{+}  \mathcal{K}(\psi_{+})}{\alpha_{-}}
            + \frac{a_3 a_4 \zeta_{-} \mathcal{K}(\psi_{-})}{\alpha_{+}}
            - \frac{a_4 c_1 \zeta_{+} \mathcal{E}(\psi_{+})}{\alpha_{-}}
            - \frac{a_3 c_2 \zeta_{-} \mathcal{E}(\psi_{-})}{\alpha_{+}}
            \right)
            \frac{Q_1}{\rho^{2}}
            \\
            &+
            \left(
            \frac{a_3 c_4 \mathcal{E}(\psi_{-})}{\alpha_{+}}
            - \frac{a_4 c_3 \mathcal{E}(\psi_{+})}{\alpha_{-}}
            + \frac{a_3 a_4 \mathcal{K}(\psi_{+})}{\alpha_{-}}
            - \frac{a_3 a_4 \mathcal{K}(\psi_{-})}{\alpha_{+}}
            \right)
            \frac{Q_2}{\rho_{+}}
            \Biggr]
    \end{split}
    \label{Eq:FzLongitudinal}
\end{equation}
with two auxiliary functions $Q_1$ and $Q_2$ based on the elliptic integrals
\begin{align*}
    Q_1 (\alpha_{+}, \alpha_{-}, \psi_{+}, \psi_{-}, a_1, a_2, c_1, c_2)                               & = \frac{a_2 \mathcal{E}(\psi_{-})}{ \alpha_{+}}
    - \frac{a_1 \mathcal{E}(\psi_{+})}{\alpha_{-}}
    + \frac{c_1 \mathcal{K}(\psi_{+})}{\alpha_{-}}
    - \frac{c_2 \mathcal{K}(\psi_{-})}{\alpha_{+}},                                                                                                                    \\
    Q_2 (\alpha_{+}, \alpha_{-}, \psi_{+}, \psi_{-}, \rho_{+}, \rho_{-}, \zeta_{+}, \zeta_{-}, \beta ) & = \frac{\rho_{+} \zeta_{+} \mathcal{K}(\psi_{+})}{\alpha_{-}}
    + \frac{\rho_{+} \zeta_{-} \mathcal{K}(\psi_{-})}{\alpha_{+}}
    + \frac{\rho_{-} \zeta_{+} \mathcal{\Pi}(\beta, \psi_{+})}{\alpha_{-}}
    + \frac{\rho_{-} \zeta_{-} \mathcal{\Pi}(\beta, \psi_{-})}{\alpha_{+}}.
\end{align*}
and the following auxiliary variables\footnotemark{}
\begin{alignat*}{4}
     & \rho_{\pm}   = R\magn \pm \rho, \quad\quad
     &                                            & \zeta_{\pm}  = \frac{L\magn}{2} \pm z, \quad\quad\quad
     &                                            & \beta = \frac{4 \rho R\magn}{\rho_{+}^2} \quad\quad\quad
     &                                            & \phantom{a}                                              \\
     & a_1          = \rho_{+}^2 + \zeta_{+}^2,
     &                                            & a_2         = \rho_{+}^2 + \zeta_{-}^2,
     &                                            & a_3 = \rho_{-}^2 + \zeta_{+}^2,
     &                                            & a_4 = \rho_{-}^2 + \zeta_{-}^2,                          \\
     & \alpha_{+}  = \frac{1}{\sqrt{a_1}},
     &                                            & \alpha_{-}  = \frac{1}{\sqrt{a_2}},
     &                                            & \psi_{+} = \frac{4 \rho R\magn}{a_1},
     &                                            & \psi_{-} = \frac{4 \rho R\magn}{a_2},                    \\
     & b_1          = \zeta_{+}^2 + R\magn^2,
     &                                            & b_2          = \zeta_{-}^2 + R\magn^2,
     &                                            & b_3 = \zeta_{+}^2 - R\magn^2,
     &                                            & b_4 = \zeta_{-}^2 - R\magn^2,                            \\
     & c_1          = b_1 + \rho^2,
     &                                            & c_2          = b_2 + \rho^2,
     &                                            & c_3 = b_3 + \rho^2,
     &                                            & c_4 = b_4 + \rho^2.
\end{alignat*}
\footnotetext{%
    Note that \cref{Eq:FrhoLongitudinal,Eq:FzLongitudinal} are undefined at $\rho = 0$ and $\rho = \pm R\magn$.
    Outside the magnet, these singularities are removable and $\vct{F}\magn$ is extendable. %
}
The coordinate transformations from cylindrical coordinates to cartesian coordinates are given by
\begin{align}
    F_x (x,y,z)
     & =
    F_\rho(\rho, z) \cos (\varphi) \\
    F_y (x,y,z)
     & =
    F_\rho(\rho, z) \sin (\varphi) \\
    F_z (x,y,z)
     & =
    F_z (\rho, z)
\end{align}
with $\rho = \sqrt{x^2 + y^2}$ and $\varphi = \arctan \left(\frac{x}{y}\right)$\footnotemark{}.

\cref{fig:ExampleMagnet}B shows the magnetic force $\vct{F}\magn$ for a cylindrical magnet with radius $R\magn = \SI{2}{\milli\meter}$ and length $L\magn = \SI{7}{\milli\meter}$.
Calculating the magnetic force is only meaningful outside the magnet.
The magnetic force is on the order of \si{\pico\newton}, similar to the order of magnitude estimated in \cite{Palovics2022} for a similar configuration.
\footnotetext{%
    Most programming languages provide a function \texttt{arctan2(y,x)} which is defined for all $x,y \in \mathbb{R}$ and returns the correct angle $\varphi$ with respect to the quadrant of the point $(x,y)$.%
}

We also provide a Python implementation of the analytical expressions for the magnetic field and force \cite{Wirthl2023b}.

\section{Numerical examples and discussion}
\label{Sec:NumericalExamples}

In the following, we present and discuss numerical examples to demonstrate the capabilities of the proposed model.
In \cref{sec:ResultsMobilityFunctions}, we start with a two-dimensional example where we investigate the influence of the mobility tensor field on the nanoparticle capture at the impenetrable wall.
Next, in \cref{sec:ResultsCylindricalMagnets}, we investigate the nanoparticle distribution in three dimensions for different positions and orientations of a finite-length cylindrical magnet, leveraging the analytic expression for the magnetic force, which we derived.
Finally, in \cref{sec:InterParticleForces}, using the analytical expressions for the magnetic field and force, we examine the validity of the assumption that the inter-particle forces are negligible compared to the magnetic force exerted by the external magnetic field.

\subsection{Influence of the mobility tensor field}
\label{sec:ResultsMobilityFunctions}

We first present a two-dimensional example where we investigate the influence of the mobility tensor field $\vct{\mathcal{M}}(\vct{x})$ on the distribution of the magnetic particles.

\begin{figure}[btp]
    \centering
    \includegraphics[width=\textwidth]{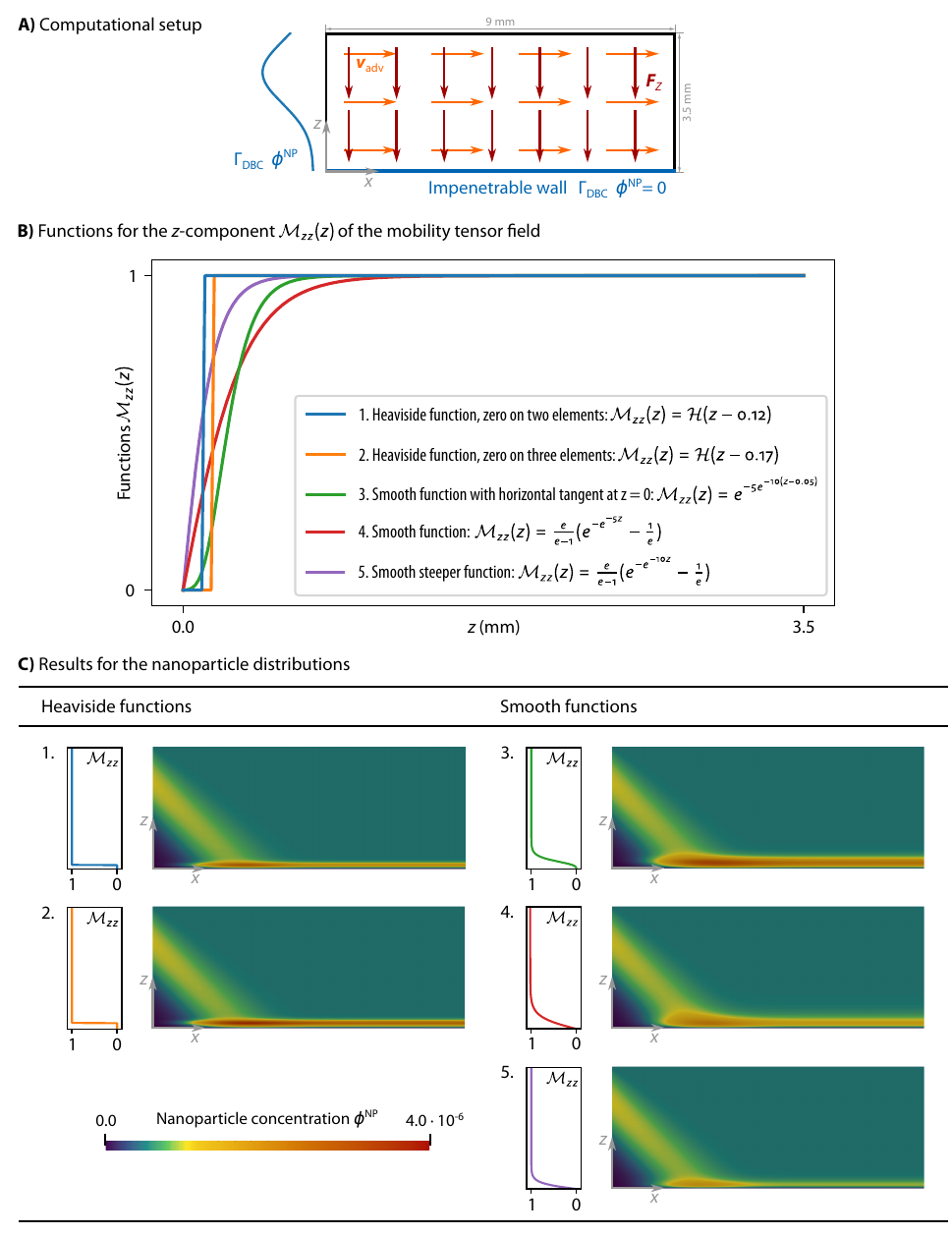}
    \caption{Investigation of the influence of the mobility tensor field on the nanoparticle distribution
        A) Computational setup.
        B) Functions for the $z$-component $\mathcal{M}_{zz}(z)$ of the mobility tensor field.
        C) Results for the nanoparticle distributions.
        The colourbar applies to all plots.}
    \label{fig:ResultsMobilityFunctions}
\end{figure}

The computational setup is depicted in \cref{fig:ResultsMobilityFunctions}A.
We study a two-dimensional slice in the XZ-plane with a size of $\SI{9}{\milli\meter} \times \SI{3.5}{\milli\meter}$, which is discretised with $180 \times 70$ linear rectangular elements.
The time step size is $\Delta t = \SI{1}{\second}$ and the total simulation time is $\SI{150}{\second}$.
For simplicity, we consider a constant advective flow velocity $v_{\textsf{adv}} = \SI{0.1}{\milli\meter\per\second}$ along the $x$-axis and a constant magnetic force $F\magn = \SI{0.2}{\pico\newton}$ along the $z$-axis, which is a reasonable order of magnitude for the considered cylindrical magnets (see \cref{fig:ExampleMagnet}B).
For a water-like fluid with a viscosity of $\mu^\ell = \SI{1e-3}{\pascal\second}$ and nanoparticles with a radius of $R\NP = \SI{100}{\nano\meter}$, this corresponds to a magnetophoretic velocity of $v\magn \approx \SI{0.1}{\milli\meter\per\second}$.
We assume a diffusion coefficient of $D = \SI{3e-3}{\milli\meter\squared\per\second}$.
On the inflow boundary at $x = 0$, we prescribe the concentration of nanoparticles as a Dirichlet boundary condition given by a bell-shaped function with a maximum value of $\phiNP = \num{1.0e-6}$.

The wall at the bottom of the domain is impenetrable, and we prescribe a Dirichlet boundary condition for the concentration of nanoparticles given by $\phi_{\textsf{DBC}}^{\textsf{NP}} = 0$.
Additionally, the z-component of the mobility tensor field is zero at the bottom wall, i.e., $\mathcal{M}_{zz} = 0$.
We compare the results for the nanoparticle distribution given different functions for $\mathcal{M}_{zz}(z)$, as given in \cref{fig:ResultsMobilityFunctions}B.
On the one hand, we consider the Heaviside function $\mathcal{M}_{zz}(z) = \mathcal{H}(z - \delta)$, with $\delta$ being the boundary layer thickness:
this means that the mobility of the nanoparticles is zero in the boundary layer.
We choose $\delta$ so that the boundary layer is two or three elements wide (given an element size of \num{0.05}).
On the other hand, we consider different smooth functions for $\mathcal{M}_{zz}(z)$, which have the value one inside the domain and have different slopes towards the boundary.

\cref{fig:ResultsMobilityFunctions}C presents the results given the different functions for $\mathcal{M}_{zz}(z)$.
In all cases, the nanoparticles accumulate at the impenetrable wall at the bottom of the domain, which was the primary motivation for introducing the mobility tensor field.
All functions lead to a similar distribution of the nanoparticles, with the thickness of the layer of captured nanoparticles depending on the function $\mathcal{M}_{zz}(z)$.
However, it shall be noted that the smooth functions are---as to be expected---numerically better behaved than the Heaviside function, which can cause convergence issues.

Defining a tensor field $\vct{\mathcal{M}}(\vct{x})$ is a simple way to model the accumulation of nanoparticles at an impenetrable wall.
It is worth noting that most similar studies in the literature, e.g., \cite{Furlani2006, Furlani2007, Roa-Barrantes2022}, do not clarify and also seem to not use appropriate boundary conditions for the nanoparticles at the wall.
This allows for studying the trajectories of the nanoparticles in the bulk of the fluid, but it is impossible to investigate the capture of the nanoparticles at a wall.
Only \citeauthor{Khashan2011}~\cite{Khashan2011} presented and discussed an approach for an impermeability condition at the wall:
they set the combined advective-diffusive flux to zero
\begin{equation}
    \phiNP \left(\vct{v}_{\textsf{adv}} + \vct{v}\magn \right)
    \cdot
    \vct{n}
    -
    D \vct{\nabla}\phiNP \cdot \vct{n} = 0.
    \label{eq:Khashan2011}
\end{equation}
We drop the advective velocity because any physically plausible velocity field cannot have a component perpendicular to an impermeable wall, either by directly imposing a physically plausible velocity field (as we do here) or by prescribing a no-slip boundary condition and solving the fluid equations.
\citeauthor{Khashan2011}~\cite{Khashan2011} subsequently set the normal component of the magnetophoretic velocity at the wall also to zero.
\cref{eq:Khashan2011} then reduces to the classical Neumann boundary condition $D \vct{\nabla}\phiNP \cdot \vct{n} = 0$, which we also impose.
In sum, their boundary condition is equivalent to our approach based on setting the normal component of the mobility tensor to zero, i.e., $\mathcal{M}_{zz} = 0$.

Nevertheless, \citeauthor{Khashan2011}~\cite{Khashan2011} also stated that their employed boundary condition poses a numerical challenge due to the steep concentration gradient at the wall.
They solve this problem by prior grid refinement adaptive to the magnetic field gradient.
We circumvent this problem by setting the mobility to zero on several elements or by using a smooth function.

\subsection{Nanoparticle capture with a cylindrical magnet of finite length}
\label{sec:ResultsCylindricalMagnets}

We now investigate a three-dimensional example with a cylindrical magnet positioned below the fluid domain.
The analytical solution for the magnetic force enables us to efficiently compare different orientations of a cylindrical magnet.

The computational setup is the one sketched in \cref{fig:IdealisedSetup}.
The domain has a size of $\SI{9}{\milli\meter} \times \SI{4}{\milli\meter} \times \SI{3.5}{\milli\meter}$, which is discretised with $180 \times 160 \times 70$ linear hexahedral elements.
The time step size is again $\Delta t = \SI{1}{\second}$ and the total simulated time $\SI{150}{\second}$.
For simplicity, we also again assume a constant advective flow velocity of $v_{\textsf{adv}} = \SI{0.1}{\milli\meter\per\second}$.
The parameters for the magnetic nanoparticles, the magnet, and the fluid are given in \cref{tab:ParametersMagneticForce}.
The nanoparticle concentration on the inflow boundary is again defined by \cref{Eq:ResultsMobilityFunctions:DBC} and zero at the bottom wall.
We use a smooth function for the mobility tensor field, i.e., Function 4 shown in \cref{fig:ResultsMobilityFunctions}B and discussed in the previous subsection.
\begin{table}[tbp]
    \centering
    \caption{Parameters for the magnetic nanoparticles, the magnet and the fluid}
    \label{tab:ParametersMagneticForce}
    \begin{tabular}{ lllll }
        \toprule
        \textbf{Symbol}                                     & \textbf{Parameter}          & \textbf{Value} & \textbf{Units}                       & \textbf{Ref.}      \\
        \midrule
        \multicolumn{2}{l}{\textit{Magnetic nanoparticles}} &                             &                &                                                           \\
        $R\NP$                                              & Radius of the nanoparticles & \num{100}      & \si{\nano\meter}                     & \cite{Furlani2006} \\
        $D$                                                 & Diffusion coefficient       & \num{3e-3}     & \si{\milli\meter\squared\per\second} & Assumed            \\
        $M_{\textsf{sp}}$                                   & Saturation magnetisation    & \num{478}      & \si{\kilo\ampere\per\meter}          & \cite{Furlani2006} \\
        \midrule
        \multicolumn{2}{l}{\textit{Magnet}}                 &                             &                &                                                           \\
        $R\magn$                                            & Radius of the magnet        & \num{2.5}      & \si{\milli\meter}                    & Assumed            \\
        $L\magn$                                            & Length of the magnet        & \num{5.0}      & \si{\milli\meter}                    & Assumed            \\
        $M_s$                                               & Magnetisation of the magnet & \num{1e6}      & \si{\ampere\per\meter}               & \cite{Furlani2006} \\
        \midrule
        \multicolumn{2}{l}{\textit{Fluid (water)}}          &                             &                &                                                           \\
        $\mu^\ell$                                          & Dynamic viscosity           & \num{1e-3}     & \si{\pascal\second}                  & Known              \\
        \bottomrule
    \end{tabular}
\end{table}
The cylindrical magnet has a radius of $R\magn = \SI{2.5}{\milli\meter}$ and a length of $L\magn = \SI{5.0}{\milli\meter}$ and is centered below the domain with a distance of $\SI{0.2}{\milli\meter}$ to the bottom wall.
In the first step, we compare three different orientations of the magnet:
A) The magnet is oriented vertically (along the z-axis);
B) The magnet is oriented horizontally (along the x-axis);
C) The magnet is rotated \SI{45}{\degree} around the y-axis.

\begin{figure}[tbp]
    \centering
    \includegraphics[width=\textwidth]{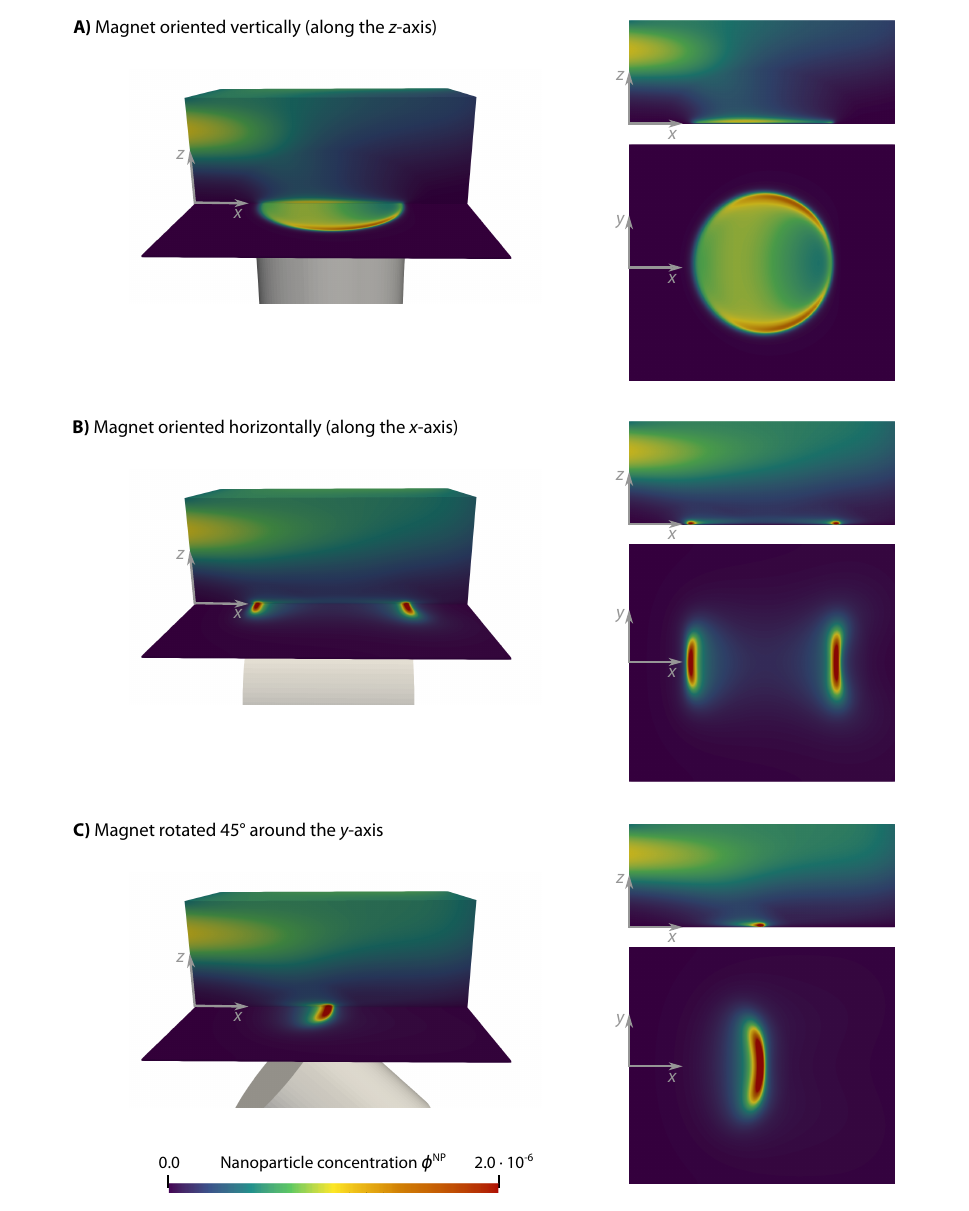}
    \caption{Results for the nanoparticle capture with a cylindrical magnet of finite length positioned below the domain.
        The colourbar applies to all plots.
    }
    \label{fig:ResultsCylindricalMagnets}
\end{figure}

\cref{fig:ResultsCylindricalMagnets} shows the concentration of the nanoparticles for the three different orientations of the magnet.
For the vertical orientation, the nanoparticles are attracted to the magnet and accumulate at the bottom wall in a circular shape directly above the magnet, similar to experimental results, e.g., \cite{Kappes2022}.
For the horizontal orientation, the nanoparticles accumulate above the two ends of the magnet, forming two ellipses.
For the \SI{45}{\degree} orientation, the nanoparticles form one ellipse above where the edge of the magnet is closest to the bottom wall.
Examples in the literature are restricted to a single orientation of a cylindrical magnet of infinite length, e.g., \cite{Furlani2006, Hewlin2023}.
In particular, we show that the nanoparticles accumulate above the ends of the magnet---which can obviously not be investigated with a magnet of infinite length.

Further, we here leverage what \citeauthor{Derby2010}~\cite{Derby2010} and \citeauthor{Caciagli2018}~\cite{Caciagli2018} stated:
their derived analytical expressions for the magnetic field of magnetised cylinders are especially convenient for applications where magnetic forces on magnetic dipoles are required---nanoparticles being one such example.
Nevertheless, our results for the magnetic force are restricted to a cylindrical magnet of finite length with longitudinal magnetisation.
Similar analytical solutions for cylindrical magnets with arbitrary magnetisation can also be derived based on the respective analytical expressions for the magnetic field presented by \citeauthor{Caciagli2018}~\cite{Caciagli2018}.
However, if the magnet is of an arbitrary shape, the magnetic field and force must be evaluated based on numerically solving Maxwell's equations.

Several studies in the literature, e.g., \cite{Furlani2006, Furlani2007, Etgar2010, Shaw2018, Yeo2021, Hewlin2023} reduced the setup to a two-dimensional problem in the XZ-plane and assumed the cylindrical magnet to be infinitely long.
In this case, the magnetic force can also be expressed analytically, as derived in \cite{Furlani2006}, and given by
\begin{align}
    F_x & = - \mu_0 V\NP f(|\vct{H}|) M_s^2 R\magn^4
    \frac{x}{2 \left( x^2 + z^2 \right)^3}           \\
    F_z & = - \mu_0 V\NP f(|\vct{H}|) M_s^2 R\magn^4
    \frac{z}{2 \left( x^2 + z^2 \right)^3},
\end{align}
where the coordinate system is at the centre of the magnet, and the longitudinal axis of the magnet is perpendicular to the XZ-plane.
We now compare results based on this assumption of an infinitely long magnet to the results for a finite-length magnet, as derived in this contribution.
In both cases, we assume that the magnet has a radius of $R\magn = \SI{2.5}{\milli\meter}$ and a distance of $\Delta = \SI{0.2}{\milli\meter}$ to the bottom boundary of the domain.
The cylindrical magnet of finite length has a length of $L\magn = \SI{5.0}{\milli\meter}$, as used in the previous examples.
For simplicity, we here assume that $f(|\vct{H}|) = 1.0 = \text{const}$ in both cases.

\begin{figure}[tbp]
    \centering
    \includegraphics[width=\textwidth]{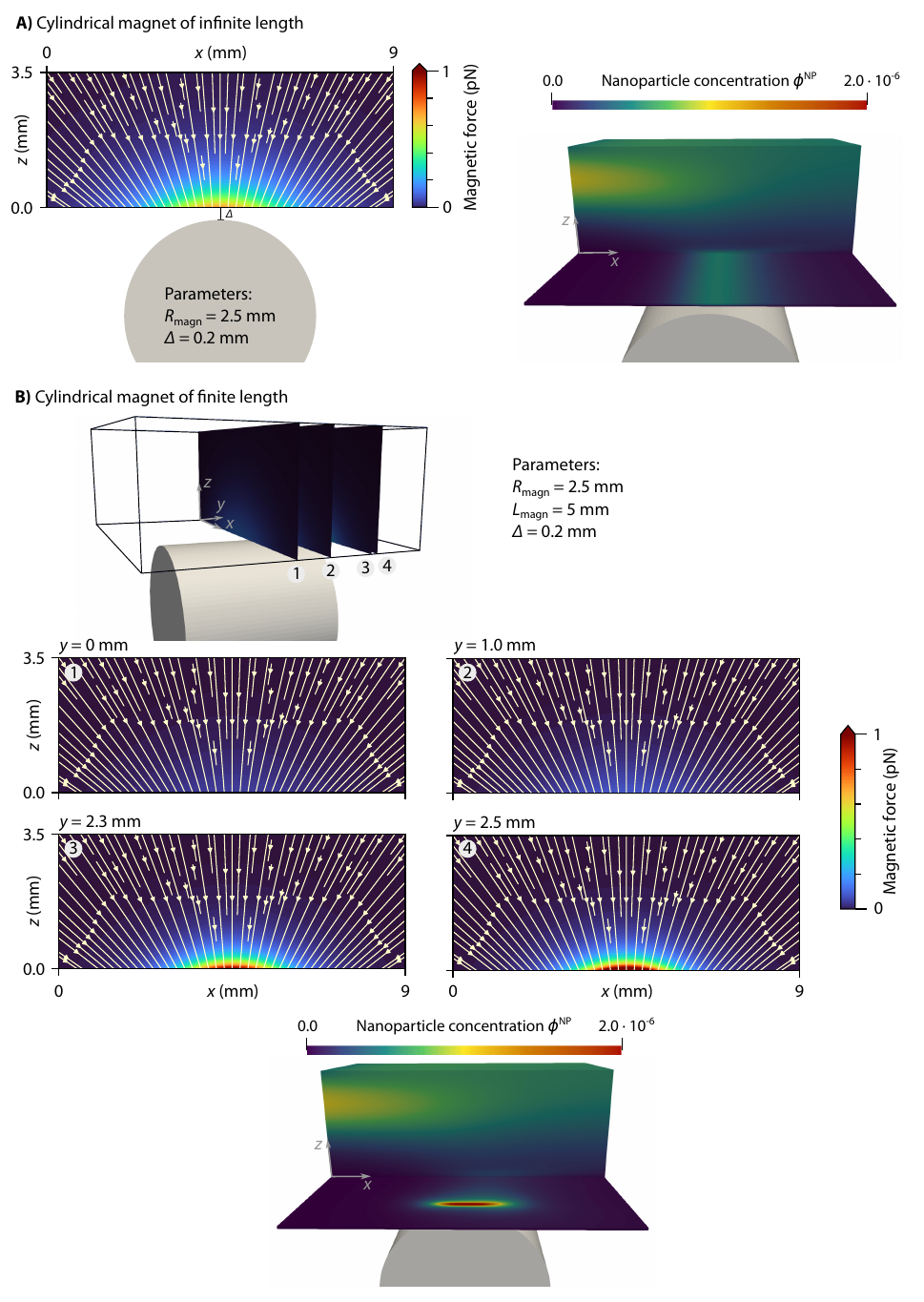}
    \caption{Comparison of the magnetic force and the resulting nanoparticle distribution for a cylindrical magnet of A) infinite length and B) finite length with $L\magn = \SI{5.0}{\milli\meter}$.
    }
    \label{fig:ResultsForceComparison}
\end{figure}

\cref{fig:ResultsForceComparison} shows the magnetic force and the resulting nanoparticle distribution for the magnet of infinite length compared to the finite-length magnet.
As is also evident in \cref{fig:ExampleMagnet}, the magnetic force of the finite-length magnet varies along the longitudinal axis of the magnet, and so we compare the force in different slices along the longitudinal axis, in this case the $y$-axis (but the $z$-axis in \cref{fig:ExampleMagnet}).
As evident in \cref{fig:ResultsForceComparison}, the direction of the magnetic force is the same in all cases, but the magnitude is significantly different.
For the cylindrical magnet of infinite length, the maximum force in the domain is \SI{0.72}{\pico\newton}.
For the finite-length magnet, the maximum force varies considerably depending on the position of the slice:
the maximum magnitude is \SI{0.06}{\pico\newton}, \SI{0.09}{\pico\newton}, \SI{1.02}{\pico\newton} and \SI{1.65}{\pico\newton} for the slices at $y = \SI{0.0}{\milli\meter}$, $\SI{1.0}{\milli\meter}$, $\SI{2.3}{\milli\meter}$ and $\SI{2.5}{\milli\meter}$, respectively.
Accordingly, the nanoparticle distributions are also markedly different:
the nanoparticles accumulate in a higher concentration above the ends of the finite-length magnet than along the infinitely long magnet.

In sum, one has to be aware that the assumption of an infinitely long magnet leads to significantly different results than a finite-length magnet.
The analytical solution for the finite-length magnet---as derived in this contribution---provides a simple and computationally efficient way to investigate the transport of nanoparticles in a more realistic setup.

\subsection{Comparison of the force exerted by the permanent magnet to the inter-particle forces}
\label{sec:InterParticleForces}

In this contribution, we only consider the external magnetic force the permanent magnet exerts on the nanoparticles.
However, the nanoparticles also exert forces on each other, and thus the question arises when these inter-particle forces are negligible compared to the force exerted by the permanent magnet.
So far in this contribution, we have assumed that the low concentration of nanoparticles ensures that the inter-particle distance is large enough for the inter-particle forces to be negligible, similar to \cite{Khashan2011, Keaveny2008, Han2010, Furlani2006, Woinska2013}.

In general, the cut-off length of dipole-dipole interactions in nanoparticle assemblies is about three particle diameters \cite{Barrera2021}.
Assuming that the nanoparticles are more than three particle diameters apart seems reasonable for the nanoparticles dissolved in the flowing fluid in our previous examples.
However, when the nanoparticles accumulate at the bottom of the domain, they come very close to each other, and thus the inter-particle forces might become relevant there.

\begin{figure}[tb]
    \centering
    \includegraphics[width=\textwidth]{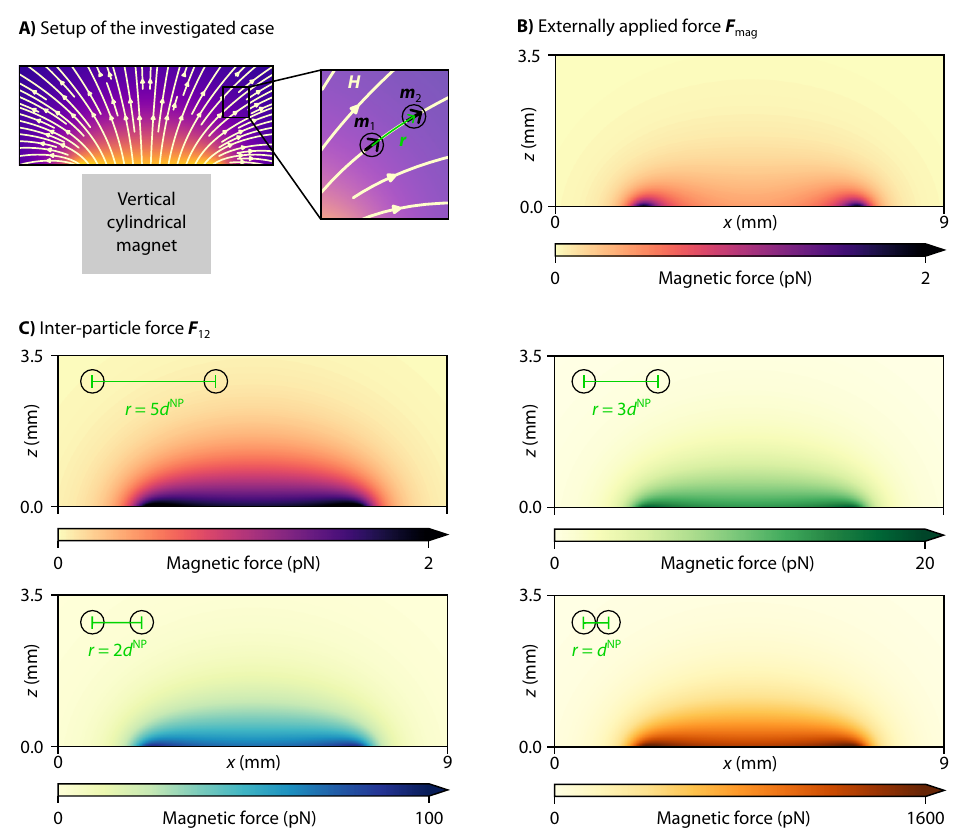}
    \caption{Comparison of the force exerted by the permanent magnet to the inter-particle forces
        A) Setup of the investigated case
        B) Externally applied force $\vct{F}\magn$
        C) Inter-particle force $\vct{F}_{12}$ between two nanoparticles with a distance $r$ between their centres.
        Note the different orders of magnitude of the forces, which are represented by the different colourmaps used in the subfigures.
    }
    \label{fig:ResultsInterParticleForces}
\end{figure}

Therefore, we compare the external magnetic force to the inter-particle forces.
We use our analytical expressions for the magnetic field and the external magnetic force and build on the force comparison presented by \citeauthor{Palovics2022}~\cite{Palovics2022}, who investigated a similar setup.
We analyse a simplified example shown in \cref{fig:ResultsInterParticleForces}A:
we consider two nanoparticles with a diameter of $d\NP = \SI{200}{\nano\meter}$ and a distance $\vct{r}$ between their centres.
The cylindrical magnet is positioned vertically below the domain (see previous example \cref{fig:ResultsCylindricalMagnets}A).
We assume that the two nanoparticles are aligned with the magnetic field $\vct{H}$ such that $\vct{r}$ is parallel to $\vct{H}$.
We again assume $f(|\vct{H}|) = 1.0 = \text{const}$ for simplicity.

In the following, the non-bold symbols denote the magnitudes of the vectors, e.g., $r = |\vct{r}|$, and a hat denotes the unit vector in the given direction, e.g., $\vct{\hat{r}} = \vct{r} / r$.

As discussed in \cref{Sec:MagneticForceOnTheNanoparticles}, the nanoparticles are modelled as point dipoles, with the magnetic moment $\vct{m}_1$ of nanoparticle \circled{1} given by
\begin{equation}
    \vct{m}_1 = V\NP \vct{H}.
    \label{Eq:MagneticMoment}
\end{equation}
Thus, the magnetic moment of the nanoparticle is aligned with the applied magnetic field.
Since the nanoparticles are much smaller than the computational domain, we assume that $\vct{H}(\vct{x}_1) = \vct{H}(\vct{x}_2)$ and hence $\vct{m}_1 = \vct{m}_2$.
The magnetised nanoparticle \circled{1} generates a magnetic field $\vct{H}_1$ at the position $\vct{r}$ of nanoparticle \circled{2} given by \cite{Jackson2021} as
\begin{equation}
    \vct{H}_1
    =
    \frac{1}{4 \pi r^3} \left[ 3(\vct{m}_1 \vct{\hat{r}}) \vct{\hat{r}} - \vct{m}_1 \right].
    \label{Eq:MagneticFieldParticle}
\end{equation}
In our case, $\vct{m}_1 \parallel \vct{r}$ and \cref{Eq:MagneticFieldParticle} simplifies to
\begin{equation}
    \vct{H}_1
    =
    \frac{1}{2 \pi r^3} \vct{m}_1.
    \label{Eq:MagneticFieldParticleSimplified}
\end{equation}
Hence, the total magnetic field $\vct{H}_2^*$ at the position $\vct{r}$ of nanoparticle \circled{2} is given by
\begin{equation}
    \vct{H}_2^*
    =
    \vct{H} + \vct{H}_1
\end{equation}
and accordingly, the magnetic moment of nanoparticle \circled{2} also changes to
\begin{equation}
    \vct{m}_2^*
    =
    V\NP \vct{H}_2^*.
\end{equation}
The magnetic moments of both particles increase due to the cross-effects.
The new values for the magnetic moments can be substituted back into the previous equations to calculate a second correction of the magnetic field and magnetic moments.
In practice, this is not necessary, and we omit it \cite{Han2010, Palovics2022}.

The force $\vct{F}_{12}$ between the two particles, i.e., the inter-particle force, is given by \cite{Griffiths2017} as
\begin{equation}
    \vct{F}_{12}
    =
    \frac{3 \mu_0 m^*_1 m^*_2}{4 \pi r^4}
    \left[
        \vct{\hat{r}} \left( \vct{\hat{m}}^*_1 \vct{\hat{m}}^*_2 \right)
        + \vct{\hat{m}}^*_1 \left( \vct{\hat{r}} \vct{\hat{m}}^*_2 \right)
        + \vct{\hat{m}}^*_2 \left( \vct{\hat{r}} \vct{\hat{m}}^*_1 \right)
        - 5 \vct{\hat{r}} \left( \vct{\hat{r}} \vct{\hat{m}}^*_1 \right) \left(\vct{\hat{r}} \vct{\hat{m}}^*_2 \right)
        \right]
    =
    - \frac{3 \mu_0 m^*_1 m^*_2}{2 \pi r^4} \vct{\hat{r}}.
    \label{Eq:ForceParticle}
\end{equation}

We evaluate the inter-particle force $\vct{F}_{12}$ for different distances $r$ between the two nanoparticles:
$r \in \bigl\{5 d\NP, 3 d\NP,\allowbreak 2 d\NP, d\NP \bigr\}$.
\cref{fig:ResultsInterParticleForces}B shows the force $\vct{F}\magn$ exerted by the external magnet and \cref{fig:ResultsInterParticleForces}C the inter-particle force $\vct{F}_{12}$.
For a distance of five particle diameters, the forces are on the same order of magnitude, namely \si{\pico\newton}.
However, the inter-particle force strongly increases for smaller distances:
for a distance of one particle diameter, it is about three orders of magnitude larger than the force of the external magnet, especially for the particles at the bottom of the domain.
This is in good agreement with the results of \citeauthor{Palovics2022}~\cite{Palovics2022}.

These results underline that one cannot simply assume that the inter-particle forces are negligible but must carefully assess whether they are relevant in the configuration studied with the assumptions made.

\section{Conclusion}
\label{Sec:Conclusion}

In this contribution, we presented a continuum approach based on the Smoluchowski advection-diffusion equation to model the capture of magnetic nanoparticles under the combined effect of fluid flow and magnetic forces.
We included a simple and numerically stable way to consider an impenetrable boundary where the nanoparticles are captured.
Further, the analytical expression for the magnetic force of a cylindrical magnet of finite length on the magnetic nanoparticles, which we derived, provides an efficient way to model the capture of magnetic nanoparticles in a more realistic setup in three dimensions.

Since many novel nanoparticle designs fail in clinical trials, our modelling efforts can help to gain insight into the behaviour of magnetic nanoparticles and help to design novel prototypes.
While our expression for the magnetic force is restricted to cylindrical magnets, this is the configuration that is commonly used in experiments, e.g., when studying magnetic nanoparticles in fluidic devices \cite{Nguyen2020, Behr2022, Kappes2022}.
Hence, such an \textit{in silco} model can help with experimental design to limit the number of experiments and thus the costs to the most promising configurations.
Finally, the presented model can serve as a precursor to more complex models, e.g., including magnets of arbitrary shape or considering complex biomechanical models coupling the transport of the nanoparticles in the blood vessels with the crossing of the vessel walls and the accumulation in the tumour tissue---both \textit{in vivo} and \textit{in silico} \cite{Kremheller2019, Wirthl2020, Wirthl2023c}.

\section*{Acknowledgements}
WAW was supported by BREATHE, a Horizon 2020|ERC-2020-ADG project, grant agreement No. 101021526-BREATHE.

\nolinenumbers

\printbibliography

\end{document}